\documentclass[10pt,twocolumn]{article}

\usepackage[english]{babel}
\usepackage[utf8]{inputenc}
\usepackage[T1]{fontenc}
\usepackage{graphicx}
\usepackage{amsmath}
\usepackage{amssymb}
\usepackage{multirow}
\usepackage{balance}
\usepackage{hyperref}
\usepackage{algorithm}
\usepackage{algpseudocode}
\usepackage[numbers,sort&compress]{natbib}
\usepackage{microtype}

\usepackage{array}
\usepackage{booktabs}
\usepackage{makecell} 
\usepackage{ragged2e} 
\usepackage{caption}

\usepackage{subcaption} 

\usepackage[most]{tcolorbox}

\usepackage{listings}
\lstdefinestyle{codeplain}{
  language={},                   
  basicstyle=\ttfamily\footnotesize, 
  breaklines=true,               
  breakatwhitespace=false,       
  columns=fullflexible,          
  keepspaces=true,               
  showstringspaces=false,
  linewidth=\columnwidth,        
  xleftmargin=0pt, xrightmargin=0pt,
  prebreak=\mbox{$\hookleftarrow$},  
  postbreak=\mbox{\textcolor{gray}{$\rightarrow$}}, 
  breakindent=0pt,
  aboveskip=2pt, belowskip=2pt   
}

\title{A Generalizable Framework for Building Executable Domain-Specific LLMs under Data Scarcity: Demonstration on Semiconductor TCAD Simulation}

\author{
    Di Wang\textsuperscript{1}, 
    Zhenhua Wu\textsuperscript{2}, 
    Yu Liu*\textsuperscript{1},
    Kai Chang*\textsuperscript{2},
    and Shaohua Wu\textsuperscript{1}\\
    \\
    \textsuperscript{1}Inspur Electronic Information Industry Co., Ltd, Beijing, China. \\
    \textsuperscript{2}Center for Quantum Matters, Zhejiang University, Hangzhou, China. \\
    \\
    *Correspondence to: Yu Liu (liuyubj@inspur.com) or Kai Chang (kchang@zju.edu.cn).
}

\date{}

\begin{document}

\maketitle

\begin{abstract}
Scientific and engineering verticals often suffer from data scarcity and strict executability requirements: models must generate not only fluent text, but also syntactically valid, tool-compilable scripts. 
We present a schema-first alignment framework for building compact, executable domain-specific LLMs in low-resource settings. 
The framework integrates three core components: 
(i) large-scale synthetic QA data generation from expert documentation to instill foundational domain knowledge; 
(ii) a code-centric IR$\rightarrow$DPO workflow that converts verified tool decks into interpretable intermediate representations (IR), performs equivalence-preserving diversification, and constructs preference pairs to directly optimize instruction compliance and code executability; 
and (iii) a controlled evaluation of Retrieval-Augmented Generation (RAG), showing that while RAG benefits general LLMs, it can marginally degrade the performance of already domain-aligned models.

We demonstrate the framework by instantiating \textsc{TcadGPT} for semiconductor Technology Computer-Aided Design (TCAD). 
Using 1.5M synthetic QA pairs and an IR-driven DPO dataset, \textsc{TcadGPT} attains 85.6\% semantic accuracy and an 80.0\% syntax pass rate on SDE executability tests, substantially outperforming state-of-the-art general LLMs such as GPT-4o. 
To probe portability beyond TCAD, we apply the same recipe to the open-source FEM solver \textit{Elmer}, observing consistent improvements in script-level success rates over general-purpose baselines. 
All datasets, benchmarks, and code (including P1, P2, and IR$\rightarrow$DPO) are released for reproducibility. 
Together, these results suggest that the proposed framework provides a robust and reproducible path toward executable LLMs in specialized, data-scarce professional domains.

\end{abstract}

\begin{figure}[t]
    \centering
    \includegraphics[width=0.32\linewidth]{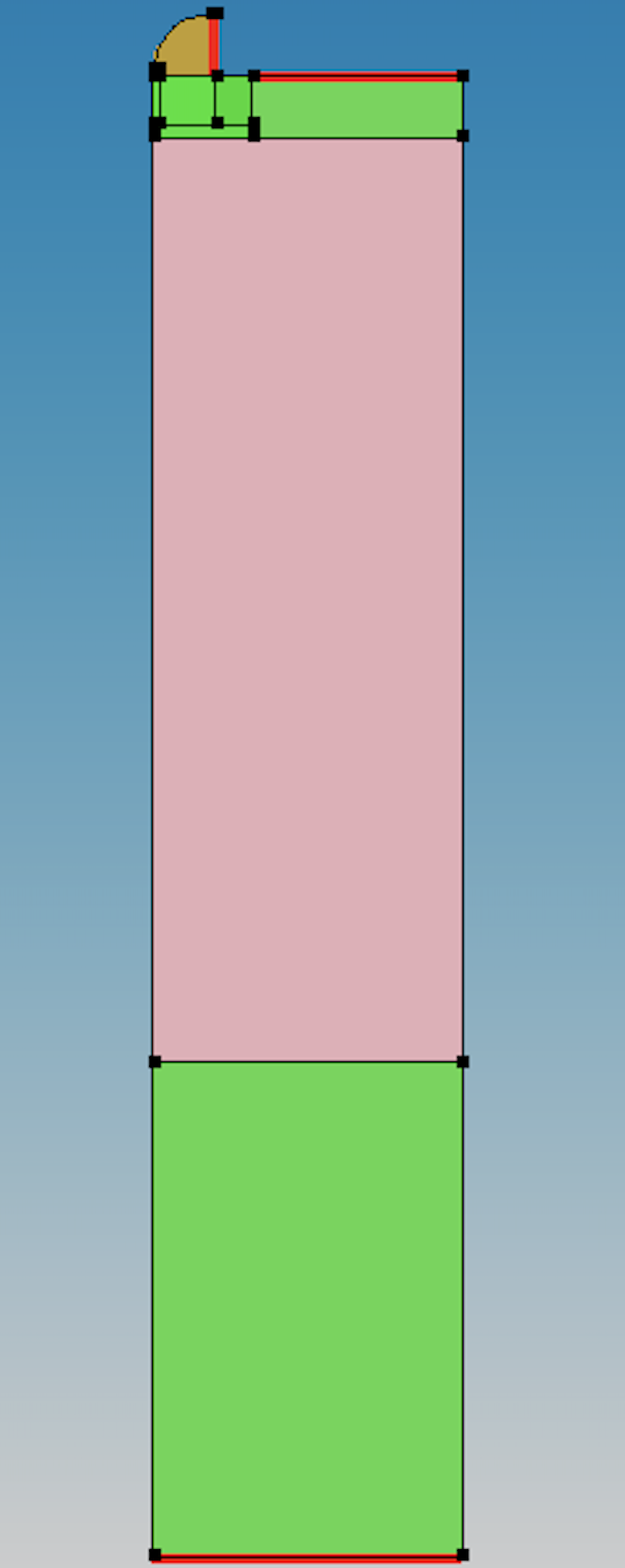}
    \caption{%
    \textbf{Executable result produced by a model-generated SDE deck.}
    A planar MOSFET structure constructed from a natural-language instruction and rendered by our model into an SDE script, then executed to completion without manual edits.
    Shown is the dopant distribution (\emph{DopingConcentration}, cm$^{-3}$) after mesh generation and export (\texttt{.tdr}/\texttt{.bnd}).
    The example includes multiple materials (Silicon, SiO$_2$, PolySi, Si$_3$N$_4$), contact definitions (gate/source/drain/substrate), region-specific doping profiles, and multi-level mesh refinement.
    This result demonstrates both \emph{instruction compliance} and \emph{tool executability}, motivating the schema-first IR$\rightarrow$DPO alignment strategy introduced later in this paper.}
    \label{fig:mosfet_example}
\end{figure}

\section{Introduction}
Advanced semiconductor research and development increasingly rely on high-fidelity digital twins that combine predictive physics with rapid, data-driven iteration. 
Within such workflows, Technology Computer-Aided Design (TCAD) serves as the computational backbone, resolving tightly coupled process–device interactions at nanometer scales and producing simulation artifacts that must be both physically accurate and \emph{tool-executable}. 
As device dimensions continue to shrink, process complexity escalates and physical effect coupling proliferates~\cite{salahuddin2018hyperscaling,cao2023future}, leading to TCAD workflows characterized by high-dimensional parameter spaces, strong nonlinearity, and script-intensive iterative setup. 
Together, these factors create an urgent demand for intelligent automation that reduces manual overhead while preserving the numerical rigor required by industrial solvers.

Recent advances in artificial intelligence (AI) have demonstrated clear potential to assist TCAD workflows, primarily through surrogate modeling and optimization. 
Neural networks have been applied to approximate electrostatic potentials, current–voltage characteristics, and other device responses, substantially reducing simulation time and computational cost~\cite{mehta2021prediction,Myung2021RestructuringTCAD,Li2022AtomisticNNProcess}. 
Physics-informed neural networks (PINNs) further integrate governing equations into learning frameworks, bridging data-driven models with physical constraints~\cite{mitusch2021hybrid,kim2023novel,lu2025physics}. 
Bayesian and Pareto-front-based optimization strategies have also shown success in reducing the number of expensive TCAD simulations required for design exploration~\cite{wu2021multiobjective,xu2022machine,jeong2024mobo}. 
However, these approaches focus primarily on accelerating numerical evaluation rather than addressing a central usability bottleneck: the authoring, debugging, and iterative refinement of complex TCAD scripts themselves. 
In industrial practice, physics-based solvers remain indispensable for accuracy, yet TCAD users continue to face steep learning curves, error-prone scripting, and limited automation across end-to-end simulation workflows.

In parallel, Large Language Models (LLMs) have rapidly advanced the state of the art in Electronic Design Automation (EDA). 
They enable natural-language interaction with design tools (e.g., ChatEDA~\cite{he2023chateda}), automate high-quality RTL and HDL generation (e.g., ChipGPT~\cite{chang2023chipgpt}, RTLLM~\cite{lu2023rtllm}), assist verification and rule checking (e.g., DRC-Coder~\cite{chang2025drc}), and support multimodal defect analysis in fabrication (e.g., FabGPT~\cite{jiang2024fabgpt}). 
Despite these successes, the application of LLMs to TCAD remains largely unexplored. 
TCAD differs fundamentally from many EDA tasks in that it requires strict adherence to simulator-specific syntax, hierarchical geometry definitions, and execution order constraints, all of which must result in \emph{tool-accepted} artifacts rather than descriptive text.

To date, the only work directly applying LLMs to TCAD is by Nguyen \textit{et al.}~\cite{sispad2024}, who fine-tuned a transformer on approximately 7{,}000 structure input files to generate TCAD scripts. 
More broadly, existing attempts typically rely on narrowly constructed datasets with limited diversity, often produced by token-level mutation of a single template. 
These studies lack standardized benchmarks and systematic comparisons with general-purpose LLMs, leaving both effectiveness and portability unclear. 
Unlike domains such as Verilog or HLS, TCAD suffers from extreme data scarcity: vendors do not release large-scale script corpora, and publicly available materials—user guides, textbooks, and bundled examples—were not designed for machine learning. 
As a result, TCAD input decks (e.g., \texttt{.cmd}, \texttt{.tdr}) pose a particularly challenging testbed for LLMs due to their nested syntax, hierarchical structure, and tight coupling between commands and physical intent.

\paragraph{Our perspective and goal.}
Rather than proposing a single-task system, we introduce a \textbf{schema-first alignment framework} for building \emph{executable} domain-specific LLMs under \emph{data scarcity}. 
We instantiate this framework on TCAD as \textsc{TcadGPT}, where the primary objective is to translate natural-language intent into simulator-accepted scripts with high instruction fidelity. 
The framework combines large-scale QA synthesis from expert documentation with a code-centric IR$\rightarrow$DPO alignment pipeline that explicitly optimizes for syntactic validity and execution order.
Figure~\ref{fig:mosfet_example} presents a representative SDE result generated by \textsc{TcadGPT} and executed without manual intervention. 
To avoid over-claiming generality from a single domain, we further include a lightweight extension study on the open-source FEM solver \textit{Elmer}~\cite{raback2012elmer}, using the same alignment recipe to probe portability beyond TCAD.

\section{A Generalizable Schema-First Alignment Framework}
\label{sec:framework}
\textbf{Design principles.} Our framework targets domains where: (i) public corpora are scarce and fragmented; (ii) solutions must yield \emph{tool-executable} artifacts (e.g., simulator scripts); and (iii) instruction-following fidelity is critical. It comprises three tightly coupled stages:
\begin{itemize}
    \item \textbf{Knowledge acquisition via QA synthesis.} Two complementary pipelines transform expert-curated documentation into \emph{Alpaca-style} instruction--response pairs at scale: \textbf{Pipeline~1} (segment-based QA) emphasizes broad coverage and phrasing robustness; \textbf{Pipeline~2} (keyword-guided QA) targets fine-grained concepts, commands, and equations for higher precision. Together they build domain priors for reasoning and syntax awareness.
    \item \textbf{Schema-first code alignment via IR$\rightarrow$DPO.} Verified executable decks are parsed into an \emph{Intermediate Representation (IR)} that preserves semantics (dimension, geometry Boolean order, contacts, doping, mesh/export). IR enables (a) \emph{equivalence-preserving diversification} and (b) \emph{deterministic instruction rendering}. We then build DPO preference pairs with single-factor, interpretable violations (numeric scale, order/procedure, export omissions) to \emph{directly optimize} instruction compliance and syntactic validity.
    \item \textbf{Retrieval as an optional, controlled component.} RAG can substantially lift generic LLMs by grounding them in domain text, but may \emph{dilute} priors for already specialized compact models. Our framework treats RAG as optional and \emph{schema-constrained}—to be used when the base is generic or when retrieval can be structured to avoid lexical overfitting.
\end{itemize}
\textbf{Outcomes.} The framework yields compact models capable of emitting \emph{tool-compilable} code and stable, high-utility QA responses. 
In the sequel, we demonstrate the framework on TCAD via \textsc{TcadGPT} and include a lightweight extension on the open-source FEM solver \textit{Elmer} to probe portability.

\section{TcadGPT: Domain-Adapted LLM for TCAD}
The rapid progress of large language models (LLMs) has enabled breakthroughs across general-purpose and scientific applications, giving rise to the emerging field of LLMs for Science (LLM4S). 
However, their applicability to highly specialized domains such as Technology Computer-Aided Design (TCAD) remains extremely limited. 
In scientific question answering, ETH Zurich introduced a multilingual LLM trained on scientific literature and Wikipedia~\cite{zthmodel}. While effective on general QA tasks, it lacks support for code understanding or procedural file generation. Other domain-specialized LLMs~\cite{jacobs2024orca,zhong2024llm4eda,liu2023chipnemo} have targeted areas like quantum chemistry, EDA scripting, or chip placement. However, these models typically rely on small-scale datasets, high-level abstraction, or prompt-only tuning. For instance, the ORCA-LLM~\cite{jacobs2024orca} focuses on generating input scripts for quantum chemistry simulations, but only covers limited domain logic with minimal instruction diversity. LLM4EDA~\cite{zhong2024llm4eda} surveys applications in hardware design, but lacks tool-specific depth. ChipNeMo~\cite{liu2023chipnemo} adapts LLMs for EDA tasks, yet does not target the code structural fidelity or low-level syntax required in TCAD environments. By comparison, our work explicitly targets both command syntax and domain semantics in a tightly scoped, low-resource setting.

This gap stems primarily from the scarcity of publicly available training data and the intricate domain-specific syntax, modeling principles, and simulation workflows required in TCAD.

Unlike other hardware design languages such as Verilog or RTL, which benefit from an active open-source ecosystem and abundant datasets via platforms like GitHub, TCAD simulation scripts are rarely shared. Most commercial TCAD tools—such as Sentaurus Process, Sentaurus Device, and Garand—are proprietary, and the input files used for real-world semiconductor simulations often contain sensitive information tied to process technology, making them strictly confidential. Consequently, only limited resources such as user guides or basic training examples are accessible for model development. This data scarcity presents a significant bottleneck in enabling LLMs to reason effectively in the TCAD domain.

As a result, existing LLMs perform poorly on TCAD-related tasks. Our preliminary experiments reveal that even state-of-the-art models such as DeepSeek V3 fail to answer code-intensive questions reliably, with limited utility in practical scenarios.
To address these challenges, we propose \textsc{TcadGPT}, a compact (8B-parameter) domain-specialized LLM tailored for semiconductor researchers working with TCAD tools.

Our work makes the following contributions:

\begin{itemize}
\item \textbf{Large-scale synthetic dataset.} We design two complementary synthetic data generation pipelines that transform publicly available TCAD manuals, textbooks, and training documents into a large-scale instruction–response corpus containing over \textbf{1.5 million} Alpaca-format samples under non-commercial academic use.
\item \textbf{Reproducible benchmark with reference answers.} 
We construct a comprehensive \textbf{264-question TCAD benchmark suite} covering physical modeling, simulation logic, and tool command syntax.
Each question is authored with an expert-provided \emph{reference answer}, and model outputs are graded using a standardized rubric that only accounts for acceptable equivalent expressions, enabling transparent and reproducible comparison across models.

\item \textbf{IR$\rightarrow$DPO alignment pipeline.} We introduce a \textbf{code-centric IR$\rightarrow$DPO alignment workflow} that extracts intermediate representations from verified TCAD decks, performs equivalence-preserving diversification, and constructs structured preference pairs with controlled instruction violations. 
This stage builds upon the QA-tuned \textsc{TcadGPT-P1\&2} model and directly optimizes instruction-following fidelity and code executability, yielding the final \textsc{TcadGPT-P1\&2+DPO} checkpoint used in Section~\ref{sec:code_exec}.
\item \textbf{Model performance and resource efficiency.} 
The QA-tuned \textsc{TcadGPT-P1\&2} model achieves a benchmark accuracy of \textbf{85.6\%} on the 264-question QA suite, significantly outperforming GPT-4o, DeepSeek V3, and R1 baselines. 
The DPO-aligned \textsc{TcadGPT-P1\&2+DPO} further demonstrates strong code executability, achieving an 80\% syntax pass rate in tool-level validation while remaining compact enough for on-premise inference on a single GPU.
\end{itemize}

Together, these contributions establish TcadGPT as a reproducible, domain-adapted foundation model for intelligent TCAD automation and digital-twin development.

\begin{figure*}[t]
    \centering
    \includegraphics[width=0.9\linewidth]{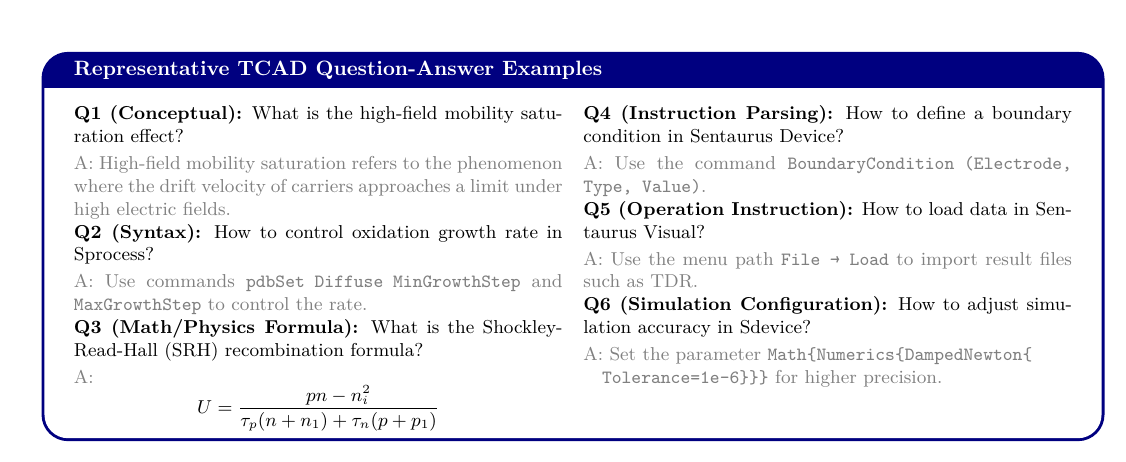}
    \caption{Representative examples from the TCAD benchmark, covering multiple question types including operational, conceptual, and model-related. These illustrate the diversity and depth of domain understanding required for high-performance models.}
    \label{fig:benchmark_examples}
\end{figure*}

\section{Experiment Setup and Benchmark}\label{sec:benchmark}

To rigorously evaluate the capabilities of \textsc{TcadGPT}, we constructed a benchmark tailored to TCAD-related tasks, designed to reflect the real-world challenges faced by semiconductor researchers. The benchmark was developed with three core principles: (1) ensuring coverage of core functional modules across TCAD toolchains, (2) including diverse question types ranging from conceptual understanding to exact syntax construction, and (3) grounding questions in realistic user scenarios. The final set consists of 264 questions carefully curated and reviewed by domain experts to ensure both technical correctness and representativeness.

\begin{figure}[htbp]
    \centering
    \includegraphics[width=0.73\linewidth]{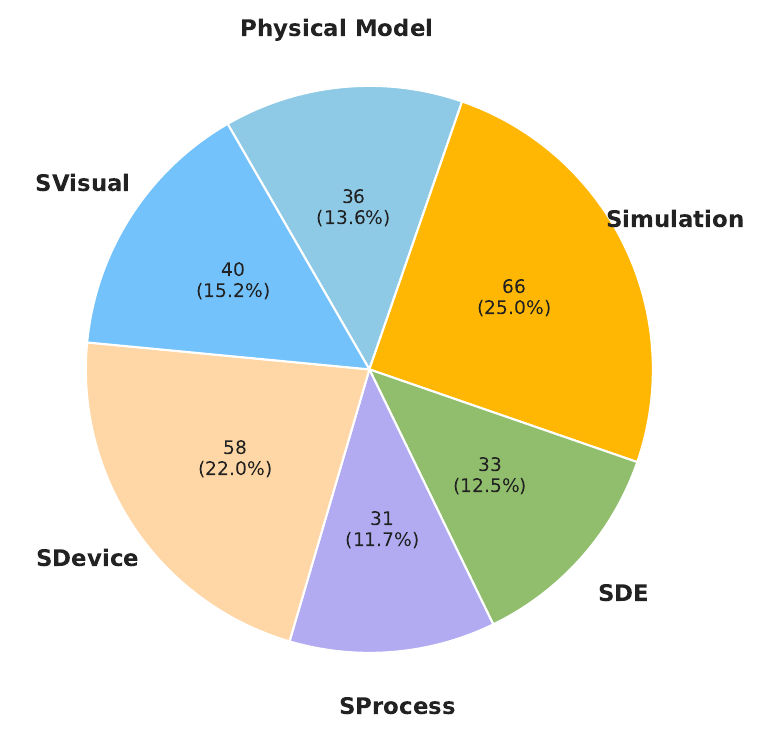}
    \caption{
        Distribution of question types in the TCAD benchmark. The six major categories—Physical Model, Simulation, SDE, SProcess, SDevice, and SVisual—cover all key simulation and modeling tasks encountered in real-world workflows.
    }
    \label{fig:question_type_dist}
\end{figure}

Figure~\ref{fig:question_type_dist} summarizes the question type distribution. The benchmark spans six major TCAD functional modules: general physical models, general simulation, \texttt{sprocess}, \texttt{sde}, \texttt{sdevice}, and \texttt{svisual}. These modules encompass both conceptual modeling knowledge and tool-specific operations, with over 65\% (173 out of 264) of the questions focusing on instruction parsing, tool operation, or syntax construction—areas identified as the most significant pain points for researchers.

All responses were evaluated against expert-provided reference answers using a standardized rubric to ensure objective and consistent scoring.
Specifically, each benchmark item is authored by a domain expert together with an unambiguous expected answer (e.g., a specific command, parameter meaning, formula, or a short canonical explanation), so that correctness can be verified without relying on subjective judgment.
The rubric is used only to handle \emph{acceptable equivalent expressions} (e.g., formatting differences, synonyms, or minor rewording) while preserving the same technical content.
For instruction-oriented questions, outputs are marked correct only if they provide directly usable commands or code snippets conforming to the expected tool syntax; for principle-oriented questions, outputs are marked correct only if they match the reference conclusion with sufficient completeness for practical TCAD use.

Representative examples from the benchmark are shown in Figure~\ref{fig:benchmark_examples}, illustrating the diversity of question types, including conceptual definitions, syntax-level commands, mathematical formulations, and simulation configurations.

\subsection*{Retrieval-Augmented Generation Setup}
To investigate whether external knowledge retrieval improves model performance, we implemented a Retrieval-Augmented Generation (RAG) pipeline using the \textit{LangFlow} platform.\footnote{\url{https://github.com/langflow-ai/langflow}} The pipeline combines offline document preparation and online query processing. Domain documents—including Synopsys Sentaurus user guides, TCAD textbooks, and official training materials—were preprocessed into manageable chunks using a custom splitter and embedded with \texttt{Ollama Embedding (nomic-embed-text)}. These embeddings were stored in \texttt{Chroma DB} for high-speed semantic search.

At inference time, user queries are first translated into English by a lightweight translation model to align with the language of the domain corpus. Retrieved document chunks are then parsed and incorporated into a structured contextual prompt, which is finally passed to the target LLM for response generation. This setup provides a consistent way to compare the standalone performance of different LLMs against their retrieval-augmented counterparts.

\subsection*{Models for Evaluation}
We evaluated two complementary aspects of model capability: (1) \textbf{QA-level understanding} and (2) \textbf{code-level executability}.  
For QA-level evaluation, we used \texttt{LLaMA 3.1 8B} as the base model and fine-tuned it using our synthetic QA corpus to obtain three variants:
\textsc{TcadGPT-P1} (trained only on data from Pipeline~1, i.e., segment-based QA generation),
\textsc{TcadGPT-P2} (trained only on data from Pipeline~2, i.e., keyword-guided QA generation),
and the combined model \textsc{TcadGPT-P1\&2} (trained on the union of both pipelines).
These variants were compared with general-purpose models including \texttt{DeepSeek V3}, \texttt{DeepSeek R1}, and \texttt{GPT-4o}.
To study the impact of retrieval augmentation, we additionally tested \texttt{LLaMA 3.1 8B + RAG}, \texttt{DeepSeek V3 + RAG}, and \textsc{TcadGPT-P1\&2 + RAG}.
This design enables a systematic comparison between (i) general-purpose versus domain-specialized LLMs, and (ii) standalone versus retrieval-augmented setups.

For code-level executability evaluation, we further assessed the \textbf{DPO-aligned model} \textsc{TcadGPT-P1\&2+DPO}, which extends the QA-tuned checkpoint with the IR$\rightarrow$DPO alignment dataset.
Unlike the QA benchmark that measures scientific correctness and descriptive reasoning, this evaluation focuses on syntactic validity and instruction-following precision, validated directly within the Sentaurus \texttt{sde -S} environment.

\paragraph{Reproducibility and Open Resources.}
To ensure transparency and reproducibility, all components of this work—including the 264-question QA benchmark, the 20-instruction SDE executability test set, the fine-tuned model weights, and the full data generation pipelines (\textbf{P1}, \textbf{P2}, and \textbf{IR$\rightarrow$DPO})—are publicly released on GitHub.\footnote{\url{https://github.com/wddddds1/TcadGPT}}
The repository provides scripts for dataset synthesis, model fine-tuning, and benchmark evaluation, enabling direct replication and further research on TCAD-oriented LLMs.

\section{QA Training Data Generation}

Due to the scarcity of publicly available TCAD-related code and documentation, we curated our training corpus from officially accessible resources such as Sentaurus User Guides, TCAD textbooks, training manuals, and bundled sample files. All documents were obtained from public sources and used under non-commercial academic conditions.

\begin{figure}[htbp]
    \centering
    \includegraphics[width=0.68\linewidth]{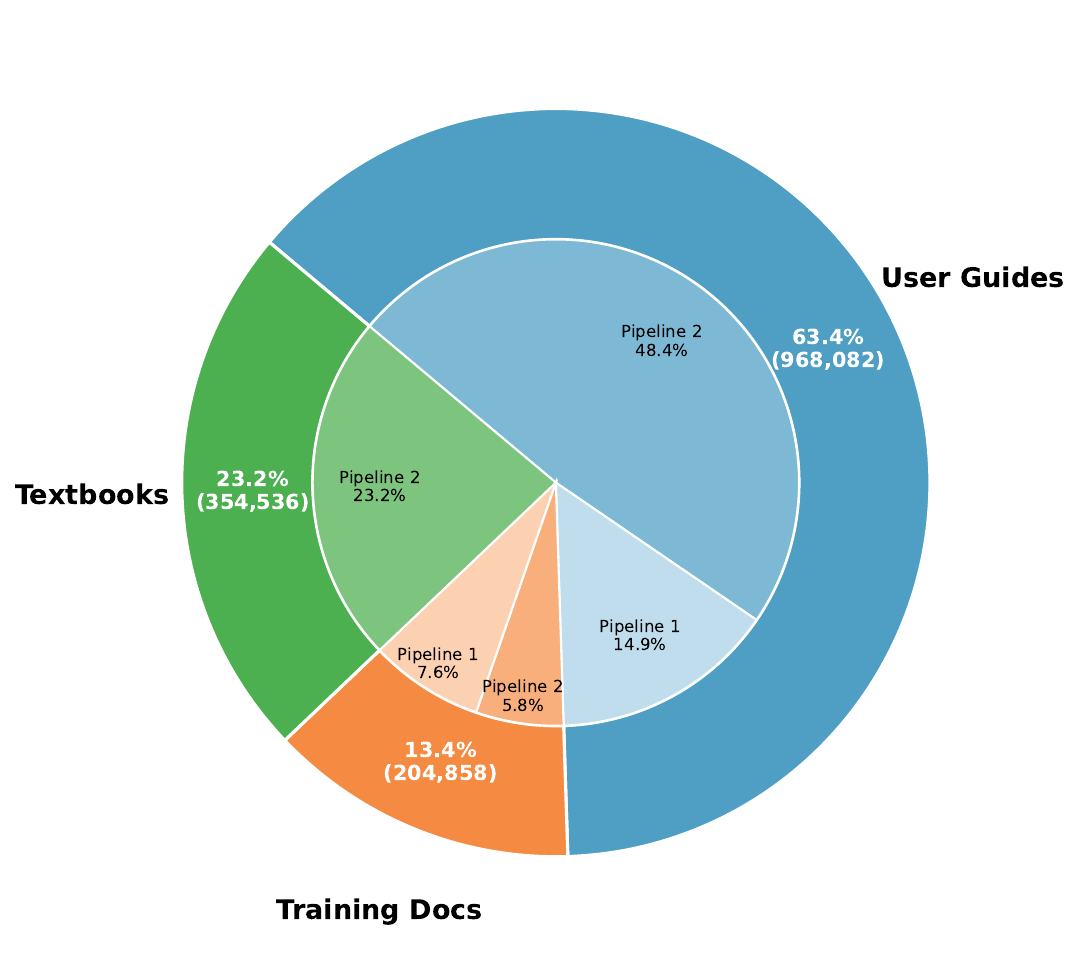}
    \caption{
        Composition of the synthetic training dataset by data source (User Guides, Training Docs, Textbooks) and data generation pipeline (Pipeline 1, Pipeline 2). The outer ring shows the total contribution of each source, while the inner ring distinguishes the two QA generation pipelines.
    }
    \label{fig:data_source_dist}
\end{figure}

To systematically construct a high-quality instruction–response dataset in Alpaca format\cite{alpaca2023}, we designed two complementary pipelines to maximize both the coverage and diversity of domain knowledge. Figure~\ref{fig:data_source_dist} visualizes the detailed breakdown of our training set, which totals over 1.5 million QA pairs. The outer ring represents the relative contributions of each document type, while the inner ring further subdivides the dataset by generation pipeline. Notably, the majority of samples originate from user guides and textbooks, with Pipeline 2 accounting for over 80\% of the total corpus.

\begin{figure}[htbp]
    \centering
    \includegraphics[width=\linewidth]{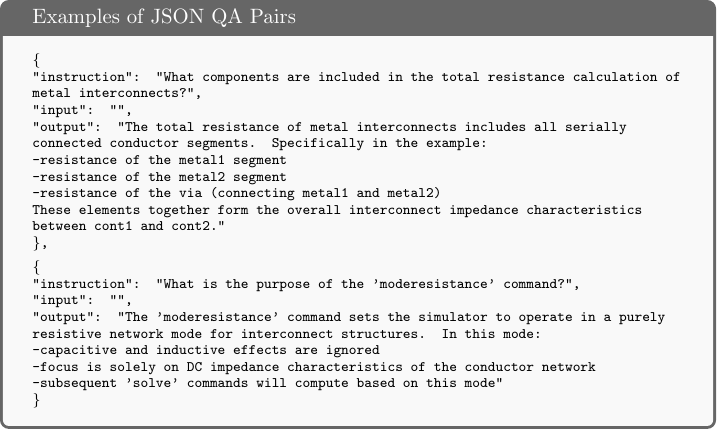}
    \caption{Examples of JSON QA Pairs used for fine-tuning.}
    \label{fig:json_examples}
\end{figure}

To illustrate the nature of the generated data, Figure~\ref{fig:json_examples} shows typical JSON-formatted QA pairs produced by our pipeline.

\paragraph{Pipeline 1: Segment-Based QA Generation}

We first converted source PDFs into text using open-source OCR tools, followed by segmentation based on structural markers such as section titles and paragraph breaks. These segments were passed to an LLM (DeepSeek V3) to generate instruction-following QA pairs. To mitigate the model's phrasing sensitivity, each question was paraphrased into ten variants using a separate LLM prompt, enhancing robustness. This process produced 340,000 samples used to train the \textsc{TcadGPT-P1} model.

\begin{algorithm}[h]
\caption{Pipeline 1: Segment-Based QA Generation}
\begin{algorithmic}[1]
\State \textbf{Input:} Set of PDF documents $\mathcal{D}$
\For{each document $d$ in $\mathcal{D}$}
    \State $T \gets \text{OCR}(d)$
    \State $S \gets \text{Segment}(T)$
    \For{each segment $s$ in $S$}
        \State $q,a \gets \text{LLM.generateQA}(s)$
        \State $Q' \gets \text{LLM.paraphrase}(q, n=10)$
        \State Add $(q,a)$ and each $(q',a)$ in $Q'$ to dataset
    \EndFor
\EndFor
\end{algorithmic}
\end{algorithm}

\paragraph{Pipeline 2: Keyword-Guided QA Generation}

To achieve greater coverage and domain precision, we designed a second pipeline that extracts technical keywords (e.g., material parameters, model names, equations) from each segment and uses them to guide QA generation. For each (segment, keyword) pair, the LLM is prompted to synthesize a relevant and focused QA instance. This method yielded over 1.2 million diverse samples and was used to train the \textsc{TcadGPT-P2} model.

\begin{algorithm}[h]
\caption{Pipeline 2: Keyword-Guided QA Generation}
\begin{algorithmic}[1]
\State \textbf{Input:} Set of PDF documents $\mathcal{D}$
\For{each document $d$ in $\mathcal{D}$}
    \State $T \gets \text{OCR}(d)$
    \State $S \gets \text{Segment}(T)$
    \For{each segment $s$ in $S$}
        \State $K \gets \text{LLM.extractKeywords}(s)$
        \For{each keyword $k$ in $K$}
            \State $q,a \gets \text{LLM.generateQA}(s, k)$
            \State Add $(q,a)$ to dataset
        \EndFor
    \EndFor
\EndFor
\end{algorithmic}
\end{algorithm}

\paragraph{Prompt Engineering and Training Setup}
To ensure the quality, coverage, and reliability of our synthetic instruction–response pairs, we designed a multi-stage prompt engineering framework. For Pipeline 1, the prompt instructed the LLM to extract all technically meaningful elements from each document segment, with an emphasis on simulation commands, modeling parameters, and domain logic. It enforced strict formatting (Alpaca-style JSON), discouraged shallow or duplicated questions, and emphasized concise, operational answers including formulas or code when applicable. To improve linguistic robustness, a separate prompt was used to generate 10 diverse paraphrases per question.

\begin{table}[t]
\centering
\caption{Overview of Prompt Types and Objectives}
\label{tab:prompt_summary}
\begin{tabular}{p{0.3\linewidth}p{0.6\linewidth}}
\toprule
\textbf{Prompt Type} & \textbf{Objective and Key Constraints} \\
\midrule
Segment QA Generation & Extract domain logic; enforce Alpaca format; emphasize command, equation, or process-level questions; maximize coverage; Chinese-only. \\
Paraphrasing for QA & Generate 10 diverse phrasings per question to improve model robustness; strictly enforce non-redundancy and valid list format. \\
Keyword Extraction & Identify all technical concepts (commands, parameters, models); enforce strict output JSON structure; ignore non-technical pages. \\
QA from Keywords & Style questions by keyword type; prohibit contextual references; ensure concise yet informative answers with examples. \\
\bottomrule
\end{tabular}
\end{table}

Pipeline 2 employed a two-step prompt structure. The first extracted all relevant keywords—such as model names, command names, or physical principles—from each segment. The second prompt used each (segment, keyword) pair to generate QA examples, with different styles based on the keyword type (e.g., definition vs. instruction). The prompt enforced structural consistency, avoided generic references (e.g., “as mentioned above”), and required detailed, example-driven answers. An overview of the prompt types and their intended objectives is summarized in Table \ref{tab:prompt_summary}.

The datasets generated from the two pipelines consisted of approximately 340,000 samples (Pipeline~1) and 1,200,000 samples (Pipeline~2). Given the limited performance of the initial 340K-sample dataset from Pipeline~1, Pipeline~2 was proposed to enhance coverage and diversity, resulting in a larger and more fine-grained corpus of 1.2M samples. Model training was conducted on 8×A100 80GB GPUs using the LLaMA 3.1 8B model and LLaMA-Factory with DeepSpeed Stage 3. Full-parameter fine-tuning was applied with a maximum sequence length of 4096 tokens, a cosine learning rate decay starting from $1\times10^{-5}$, batch size 4 per GPU with gradient accumulation of 8, and 5 total epochs. All training used \texttt{bf16} precision, with 2\% of the dataset held out for validation.

\section{Code Augmentation and DPO Dataset Construction}
\label{sec:code_dpo}

\begin{figure}[!t]
    \centering
    \includegraphics[width=\linewidth]{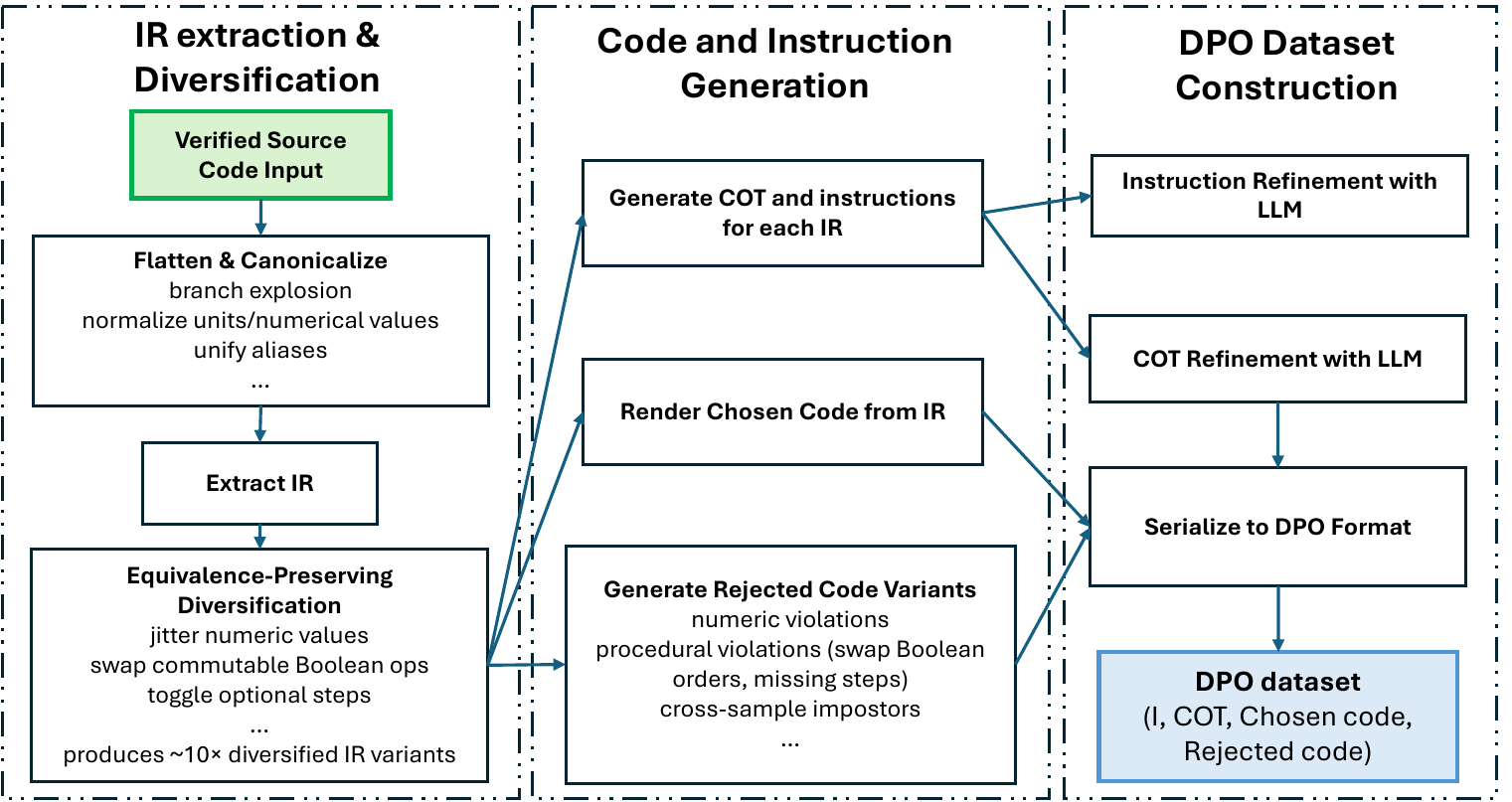}
    \caption{%
    \textbf{Overview of the IR$\rightarrow$DPO alignment pipeline.}
    Starting from verified TCAD source decks, the pipeline extracts a schema-level intermediate representation (IR), performs equivalence-preserving diversification, renders canonical instructions and reference code, and constructs preference pairs with controlled numeric and procedural violations for Direct Preference Optimization (DPO).
    }
    \label{fig:ir_dpo_pipeline}
\end{figure}

QA-style supervision provides broad domain knowledge but rarely teaches a model to \emph{follow instructions precisely} when generating executable TCAD scripts. Models trained only on QA or unstructured code often produce nearly-correct yet unusable outputs—commands in the wrong order, missing export steps, or parameters off by an order of magnitude. To bridge this gap, we introduce a \textbf{code-centric alignment pipeline} that converts a small set of verified decks into a large, interpretable dataset for Direct Preference Optimization (DPO). Figure~\ref{fig:ir_dpo_pipeline} illustrates the complete IR$\rightarrow$DPO alignment workflow, including IR extraction and diversification, instruction and code rendering, and the construction of preference pairs with controlled violations.

\subsection*{IR construction and equivalence-preserving diversification}
Starting from $\sim$800 official Sentaurus \texttt{sde}/\texttt{sprocess}/\texttt{sdevice} decks, we extract a lightweight \emph{Intermediate Representation (IR)} that captures the essential simulation facts: dimensionality and up-direction, material list, region geometry with ordered Boolean operations, contact definitions, doping specifications, and mesh/export directives. Each record also contains a compact \emph{fact card} (region counts, Boolean-order string, contact presence, expected outputs), which formalizes the invariants that must not change across transformations.

We first \emph{flatten} decks by unifying aliases, standardizing numeric formats, and canonicalizing command orders where tool semantics allow, so comparisons reflect semantics rather than style. We then perform \textbf{equivalence-preserving diversification} on the IR to enlarge coverage without altering physical meaning: mildly reordering commutable Boolean/geometry operations, applying small unit-aware numeric jitters, canonicalizing aliases, and toggling optional mesh/export statements. This yields roughly a $\times$10 expansion while remaining executable and semantically faithful.

\begin{figure}[!t]
    \centering
    \includegraphics[width=\linewidth]{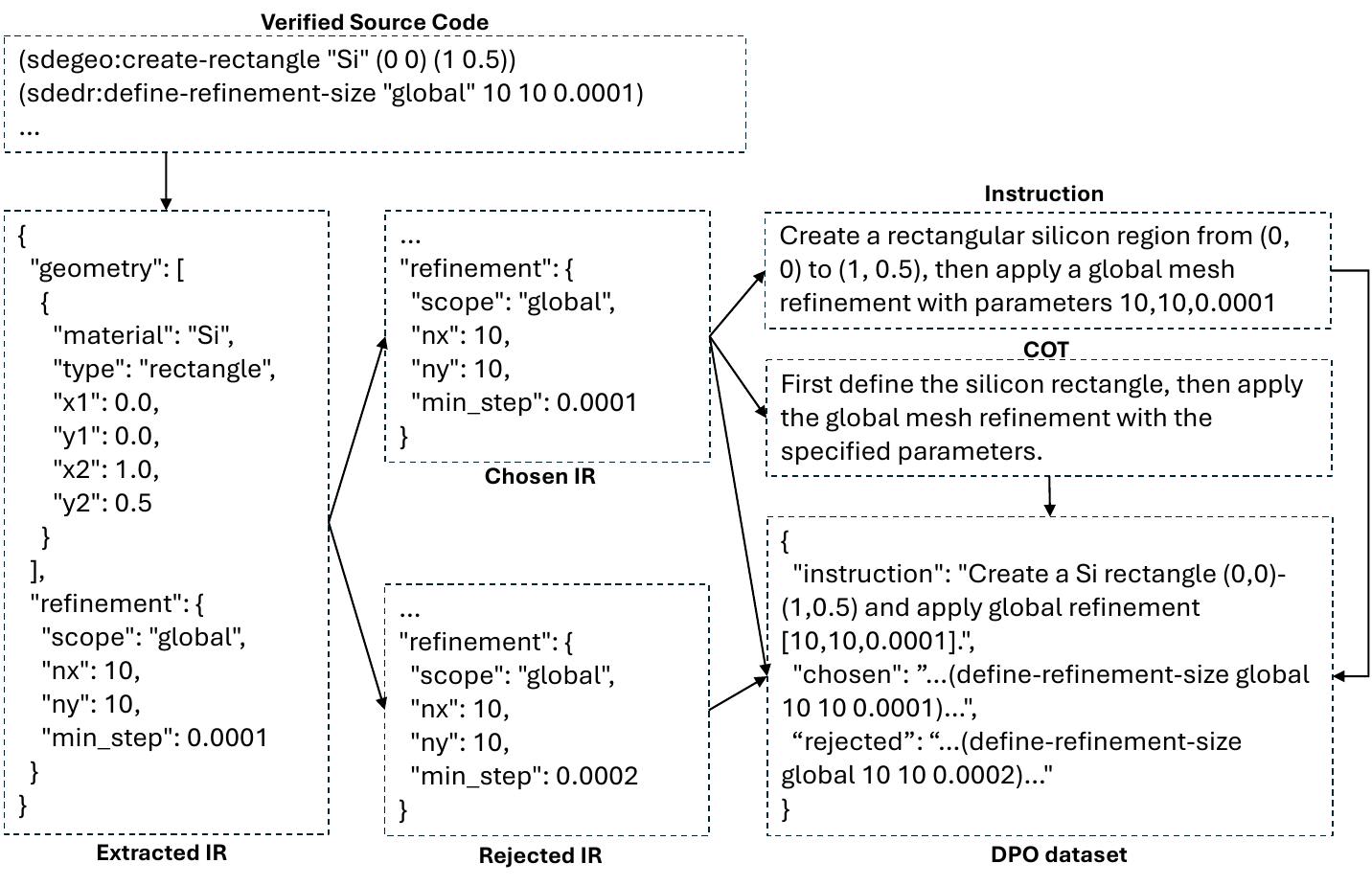}
    \caption{%
    \textbf{Example of IR-driven DPO pair construction.}
    A verified TCAD source snippet is parsed into an intermediate representation (IR), from which a canonical natural-language instruction, a chosen (instruction-compliant) code sample, and a rejected variant with a single controlled violation (e.g., a numeric mismatch) are deterministically generated.
    }
    \label{fig:ir_dpo_example}
\end{figure}

\subsection*{Instruction rendering, validation, and DPO packaging}
Each diversified IR is deterministically rendered into a \emph{canonical natural-language instruction} and a reference code snippet. A \emph{numeric whitelist} restricts visible constants to those present in the IR or reference code. To make the intended reasoning explicit without verbosity, we attach a concise, symmetric \textbf{chain-of-thought (COT)} that enumerates geometry, doping, mesh, and export steps.

To explicitly optimize instruction following, we build \textbf{DPO preference pairs}. For every instruction, the validated reference code is the \emph{chosen} sample. We synthesize several \emph{rejected} variants that intentionally violate a single interpretable dimension, including (i) \textit{numeric deviations} (near/step/mag/wide jitters or $\times10$/$\times0.1$ unit-scale errors), (ii) \textit{procedural violations} (e.g., swapping Boolean order, moving contact definitions before refinement, omitting \texttt{build-mesh} or export), and (iii) \textit{cross-sample impostors} (valid code from another record with conflicting facts). A two-stage checker validates negatives: non-subset multiset tests for numbers (with weak-accept fallbacks) and IR diffs/targeted regex for structure. 

Finally, we serialize samples into JSON with stable IDs. All pairs share the same instruction and COT—the only difference is inside the fenced code block—so the preference signal cannot be confounded by stylistic artifacts. We also produce five paraphrased instruction variants (research/engineering/concise/stepwise/conversational) filtered by the same whitelist for robustness.

\begin{algorithm}[t]
\caption{IR-to-DPO Alignment Workflow}
\label{alg:ir_dpo}
\begin{algorithmic}[1]
\State \textbf{Input:} Verified TCAD decks $\mathcal{C}$
\For{each deck $c \in \mathcal{C}$}
  \State $\mathrm{IR} \gets \textsc{ExtractIR}(c)$
  \State \hspace{1em}\emph{// dimension, geometry, doping, mesh}
  
  \State $\mathrm{IR}_{\text{flat}} \gets \textsc{Flatten}(\mathrm{IR})$
  
  \State $\{\mathrm{IR}_i\} \gets \textsc{Diversify}(\mathrm{IR}_{\text{flat}})$
  \State \hspace{1em}\emph{// order swaps, numeric jitters, toggles}
  
  \For{each $\mathrm{IR}_i$}
     \State $I \gets \textsc{RenderInstruction}(\mathrm{IR}_i)$
     \State \hspace{1em}\emph{// numeric whitelist}
     
     \State $c^{\star} \gets \textsc{RenderCode}(\mathrm{IR}_i)$
     \State \hspace{1em}\emph{// chosen code}
     
     \State $\{\tilde{c}_j\} \gets \textsc{MakeRejected}(\mathrm{IR}_i,c^{\star})$
     \State \hspace{1em}\emph{// controlled violations}
     
     \State \textsc{ValidatePairs}$(I,c^{\star},\tilde{c}_j)$
     \State \textsc{SerializeDPO}$(I,c^{\star},\tilde{c}_j)$
  \EndFor
\EndFor
\end{algorithmic}
\end{algorithm}

\paragraph*{Illustrative example.}
Figure~\ref{fig:ir_dpo_example} provides a concrete example of the IR-driven DPO construction process.
Starting from a verified TCAD source snippet, the extracted IR is used to render a natural-language instruction, a reference (chosen) code sample, and a rejected variant that differs by a single, interpretable violation.

\begin{tcolorbox}[colback=gray!2, colframe=black!40, breakable, title=Example of a DPO Pair (abbreviated)]
\label{ex:dpo_example}
\small
\RaggedRight
\textbf{Instruction:} Construct a rectangular region on a two-dimensional silicon substrate from coordinates (0, 0, 0) to (1, 1, 0). Define two global rectangular windows and place a virtual contact point at (0, 0.5, 0). Perform boron doping with a concentration of 9.8e+12 inside the windows, then refine the global mesh with parameters [10, 10, 0.0001, 1, 1, 0.0001]. Finally, build the mesh and export both BND and TDR files for use in SDevice.

\textbf{Chosen (excerpt):}
\begin{lstlisting}[style=codeplain]
(sdedr:define-refinement-size "global" 10 10 0.0001 1 1 0.0001)
...
(sdeio:save-tdr-bnd "n@node@.tdr" "n@node@.bnd")
\end{lstlisting}

\textbf{Rejected (excerpt):}
\begin{lstlisting}[style=codeplain]
(sdedr:define-refinement-size "global" 10 10 0.0001 1 1 0.0002)
...
(sdeio:save-tdr-bnd "n@node@.tdr" "n@node@.bnd")
\end{lstlisting}
\end{tcolorbox}

\section{Experiment Results and Ablation Study}\label{sec:results}

To comprehensively evaluate the effectiveness of \textsc{TcadGPT}, we designed a two-tier evaluation framework addressing both \textbf{semantic understanding} and \textbf{code executability}.

\paragraph{(1) QA-level Evaluation.}
We constructed a curated benchmark of 264 expert-verified questions spanning six key TCAD subdomains—\textbf{General Physical Model}, \textbf{Simulation}, \textbf{SDE}, \textbf{SProcess}, \textbf{SDevice}, and \textbf{SVisual}. 
This benchmark measures the model’s ability to comprehend physical principles, interpret tool syntax, and produce scientifically correct and usable responses. 
Each answer was manually scored based on scientific correctness and syntactic validity, as described in Section~\ref{sec:benchmark}.

\paragraph{(2) Code-level Executability Evaluation.}
To further assess the impact of our IR$\rightarrow$DPO alignment workflow on executable code generation, we created a separate test set consisting of 20 unseen natural-language SDE instructions automatically generated from the IR renderer.
For each instruction, the model produced one or multiple candidate scripts, which were directly validated using \texttt{sde -S} syntax checking in the official Sentaurus Structure Editor.
We report two metrics: (i) \textbf{Pass@1}, the proportion of single-sample generations passing syntax check, and (ii) \textbf{Pass@3}, the proportion passing when up to three samples are generated per instruction.
We also distinguish between \emph{direct passes} and cases requiring simple placeholder replacement (e.g., substituting \texttt{@height@} with numeric constants).

This dual-level evaluation provides complementary perspectives: the QA benchmark quantifies the model’s reasoning and instruction-following accuracy, which reflects the model's understanding of knowledge, while the SDE test set measures its ability to produce \emph{tool-valid, executable} code under realistic conditions, which reflects the application of knowledge by the model.

\begin{figure*}[t]
    \centering
    \includegraphics[width=0.95\linewidth]{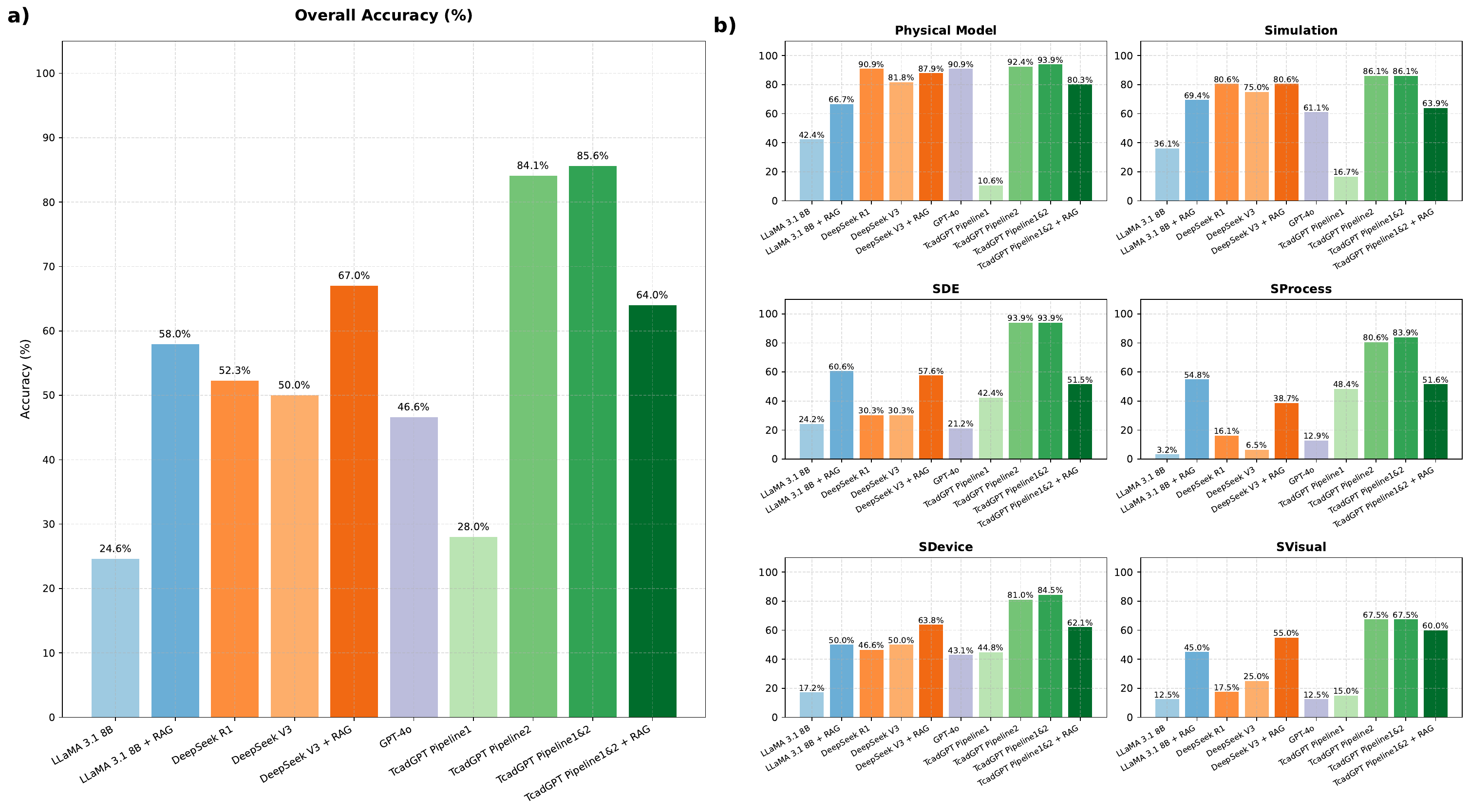}
    \caption{Model performance across all six TCAD subdomains and the overall benchmark (264 questions). Left: overall accuracy comparison across all models. Right: subdomain-level accuracy on \textbf{General Physical Model}, \textbf{Simulation}, \textbf{SDE}, \textbf{SProcess}, \textbf{SDevice}, and \textbf{SVisual}.}
    \label{fig:benchmark_result}
\end{figure*}

\subsection{QA-level Benchmark Results}
\subsubsection{Overall Performance}

Figure~\ref{fig:benchmark_result}(a) summarizes the overall accuracy achieved by each model across the entire benchmark (264 questions). Among all models, \textsc{TcadGPT-P1\&2} (trained with combined data pipelines) achieved the highest accuracy of 85.6\%, demonstrating significant superiority over general-purpose LLM baselines, including GPT-4o\cite{hurst2024gpt4o} (46.6\%), DeepSeek R1\cite{deepseek2024r1} (52.3\%), and DeepSeek V3\cite{liu2024deepseekv3} (50.0\%). Additionally, we observed substantial performance improvements through retrieval augmentation (RAG), with DeepSeek V3 improving from 50.0\% to 67.0\% and LLaMA 3.1 8B improving from 24.6\% to 58.0\%. However, notably, adding RAG to \textsc{TcadGPT-P1\&2} reduced the accuracy from 85.6\% to 64.0\%, indicating that an overly broad retrieval context may dilute domain-specific fine-tuning effectiveness.

\subsubsection{Subdomain-Level Analysis}

To gain deeper insight, we evaluated model performance on each of the six TCAD subdomains separately (Figure~\ref{fig:benchmark_result}(b)).

\textbf{General Physical Model}: \textsc{TcadGPT-P1\&2} achieved near-perfect performance (93.9\%), significantly surpassing general models like GPT-4o (90.9\%) and DeepSeek V3+RAG (87.9\%). The noticeable gap between \textsc{TcadGPT-P1} (10.6\%) and \textsc{TcadGPT-P2} (92.4\%) on the \textbf{General Physical Model} category reflects a difference in data coverage rather than optimization stability. \textsc{TcadGPT-P1} was primarily trained on segment-level extractions that were biased toward tool usage and script patterns; as a result, it frequently attempted to answer conceptual physics questions by emitting procedural instructions or simulator commands, which were graded as incorrect under our rubric. In essence, \textsc{TcadGPT-P1} became overly \emph{code-prone}---responding to general physics questions with instruction-like outputs instead of descriptive reasoning. Pipeline~2, by contrast, was generated via keyword-guided QA synthesis explicitly targeting physical mechanisms, material parameters, and modeling assumptions, which allowed it to produce high-accuracy descriptive answers rather than instruction-style code completions.

\textbf{Simulation}: \textsc{TcadGPT-P2} and \textsc{TcadGPT-P1\&2} both led performance (86.1\%), outperforming RAG-enhanced general models (DeepSeek V3+RAG at 80.6\%) and significantly exceeding baseline models such as GPT-4o (61.1\%) and LLaMA 3.1 8B (36.1\%).

\textbf{SDE}: In tasks related to mesh and geometry setups, \textsc{TcadGPT-P2} and \textsc{TcadGPT-P1\&2} achieved the best results (93.9\%), demonstrating remarkable superiority over general models, with the best general model (LLaMA 3.1 8B+RAG) reaching only 60.6\%. The sharp contrast highlights the complex, tool-specific syntax that domain-adapted models are uniquely suited to handle.

\textbf{SProcess}: Similarly, \textsc{TcadGPT-P1\&2} (83.9\%) and \textsc{TcadGPT-P2} (80.6\%) significantly outperformed all general-purpose counterparts, with the best-performing general model (LLaMA 3.1 8B+RAG) at 54.8\%. This indicates the importance of domain-specific training, particularly in complex simulation tasks such as oxidation and diffusion.

\textbf{SDevice}: \textsc{TcadGPT-P1\&2} and \textsc{TcadGPT-P2} consistently outperformed other models, achieving accuracies of 84.5\% and 81.0\%, respectively. General-purpose models struggled substantially, with GPT-4o (44.8\%) and DeepSeek V3+RAG (63.8\%) underscoring the difficulty in correctly parsing detailed device simulation commands and parameters without specialized training.

\textbf{SVisual}: Visualization tasks posed similar challenges, yet \textsc{TcadGPT-P2} and \textsc{TcadGPT-P1\&2} delivered outstanding accuracy (67.5\%), surpassing general baselines such as DeepSeek V3+RAG (55.0\%) and GPT-4o (15.0\%). The data suggests that visualization syntax, while slightly more intuitive, still greatly benefits from specialized domain data.

\subsubsection{Ablation Study}

Our ablation study compares several configurations to examine the contributions of RAG and domain-specific fine-tuning separately. General-purpose models significantly improved when supplemented with retrieval (e.g., LLaMA 3.1 8B from 24.6\% to 58.0\% and DeepSeek V3 from 50.0\% to 67.0\%), validating retrieval augmentation's effectiveness in providing necessary contextual grounding. Conversely, adding retrieval to the already specialized \textsc{TcadGPT-P1\&2} reduced overall performance from 85.6\% to 64.0\%, suggesting the carefully curated training data provided by Pipelines 1 and 2 sufficiently covers domain knowledge, and overly broad retrieved contexts may introduce noise rather than valuable insights.

Moreover, comparing \textsc{TcadGPT-P1} and \textsc{TcadGPT-P2} individually highlighted that structured keyword-based QA generation (Pipeline 2) dramatically enhanced the model's capacity to understand and correctly respond to intricate TCAD-specific queries across all modules, underscoring the importance of targeted and structured domain data.

These results collectively demonstrate that a tailored domain-specific training approach, especially structured and keyword-guided data augmentation, is vital for LLM effectiveness in highly specialized, low-resource domains like TCAD.

\paragraph{Summary}
Pipeline 2 delivers consistent and sizable gains across both conceptual and code-intensive tasks, substantially outperforming all baselines. These improvements highlight the value of targeted prompt design and keyword-guided coverage when synthesizing domain-specific QA data.

Our results further show that domain-specific fine-tuning is a stronger foundation than retrieval alone: RAG markedly lifts general-purpose models, yet naïvely adding RAG to the already specialized \textsc{TcadGPT} degrades accuracy. This suggests that unstructured retrieval can dilute learned domain priors; future work should investigate constrained or schema-aware retrieval to complement specialization without introducing noise.

\subsubsection{Error Analysis}

We conducted a qualitative analysis on failure cases of \textsc{TcadGPT-P2} to identify remaining challenges. Three main categories of errors were observed:

\textbf{(1) Invalid References:}  
Some answers contained hallucinated citations to equations or figures not present in the context. These were often inherited from training passages and incorrectly retained during generation. For example:

\begin{tcolorbox}[colback=gray!5, colframe=black!50, title=Invalid Equation Reference]
\textbf{User Query:} What are the equations corresponding to transient simulation and small-signal analysis?

\textbf{Model Answer:}  
The equation corresponding to transient simulation is \textbf{Equation 12}, which describes the basic transient formulation;  
the small-signal analysis corresponds to \textbf{Equation 48}, describing the basic principle of small-signal behavior.
\end{tcolorbox}

In this case, the model refers to "Equation 12" and "Equation 48" even though neither is defined in the input context. We introduced stricter prompts to suppress such false references, which led to modest improvements, but some cases still persist.

\textbf{(2) Command Hallucination and Misunderstanding:}  
The model sometimes generates incorrect or fabricated command syntax, particularly for less common tasks or tools with sparse training coverage. Consider the following example related to dopant definition in \texttt{Sprocess}:

\begin{tcolorbox}[colback=gray!5, colframe=black!50, title=Hallucinated Command in Sprocess]
\textbf{User Query:} How to define a new dopant species in Sprocess?

\textbf{Correct Answer:}  
Use the command \texttt{implant species=<name> <material>} and optionally specify \texttt{imp.table} to define the implant table.  
Use the \texttt{damage} flag to enable damage calculation.

\textbf{Model Answer:}  
In Sprocess, use the \texttt{dopant} command:  
\texttt{dopant add <name> <new dopant name> <species> <existing dopant name>}
\end{tcolorbox}

Here, the model invents a non-existent \texttt{dopant} command, likely due to pattern confusion or partial phrase overlap from other simulator languages.

\textbf{(3) Tokenization and OCR Noise:}  
Some errors stemmed from training data issues such as OCR artifacts, broken LaTeX equations, or split command lines. Although we applied cleanup heuristics, their effects remain localized and can interfere with parsing or token alignment.

Despite these challenges, \texttt{TcadGPT-P2} consistently outperforms baseline models in structural fidelity and domain-specific syntax. Future work may benefit from integrating simulation-aware validators, structured retrieval, or Abstract Syntax Tree (AST) level constraints to further mitigate hallucinations.

\subsubsection{Analysis of RAG-Induced Performance Degradation}
\label{sec:rag_degradation}

Although retrieval-augmented generation (RAG) is generally beneficial for open-domain tasks, we observed that adding RAG to \textsc{TcadGPT-P1\&2} resulted in a notable drop in performance. To clarify this phenomenon, we analyze a representative example from our benchmark.

\begin{tcolorbox}[colback=gray!2, colframe=black!40, title=Example (Benchmark ID 78): Question]
\small
\textbf{Question:} What are the equations corresponding to transient simulation and small-signal AC analysis, respectively?
\end{tcolorbox}

\begin{tcolorbox}[colback=gray!2, colframe=black!40, title=\textsc{TcadGPT-P1\&2}]
\small
Transient simulation uses a time-dependent differential equation describing the device’s dynamic behavior, in the form
\[
\frac{d}{dt}q(t) + f(t) = 0;
\]
small-signal AC analysis adopts a frequency-domain linearized model, such as the admittance matrix
\[
Y = A + j\omega C,
\]
which is used to compute device responses under small-signal excitation.
\end{tcolorbox}

\begin{tcolorbox}[colback=gray!2, colframe=black!40, title=\textsc{TcadGPT-P1\&2+RAG}]
\small
In the given context, transient simulation corresponds to equation~36, and small-signal AC analysis corresponds to equation~38.
\end{tcolorbox}

\begin{tcolorbox}[colback=gray!2, colframe=black!40, title=DeepSeek V3 + RAG (for comparison)]
\small
Based on the provided context, transient and small-signal AC analyses correspond to the following formulas.

\textbf{Small-Signal AC Analysis:}
The admittance matrix is described as
\[
\delta I = Y\,\delta V \quad (\text{Eq.~22}),
\]
where $\delta V$ and $\delta I$ are the complex voltage and current vectors at selected nodes.
\newline
...
\end{tcolorbox}

This comparison clearly illustrates why RAG may degrade the performance of a compact domain-specific model. 
The retrieved documents included literal equation indices (e.g., ``Eq.~36'', ``Eq.~38'') from TCAD manuals, which misled the 8B model into \emph{copying numeric references} instead of reasoning about the physical meaning of the equations. 
Because RAG directly injects unfiltered text segments without semantic abstraction, a smaller model lacking strong reasoning capacity tends to \emph{overfit to lexical cues} rather than synthesizing physically valid expressions. 

In contrast, the standalone fine-tuned \textsc{TcadGPT-P1\&2} relies on its aligned internal knowledge and produces consistent, interpretable, and executable formulas. 
This case demonstrates that while RAG benefits large general models (e.g., DeepSeek V3) through contextual recall and reasoning, it can be detrimental to lightweight domain-specific models, where noisy or contextually inconsistent retrievals may impair both \emph{instruction compliance} and \emph{code executability}.

\subsection{Code-level Executability Results}
\label{sec:code_exec}

While the QA-level benchmark evaluates semantic and reasoning accuracy, it does not directly reflect whether model-generated scripts are \emph{tool-executable}. 
To quantify the practical impact of the IR$\rightarrow$DPO alignment pipeline, we conducted a syntax-level executability test on 20 unseen SDE instructions rendered from the IR generator. 
Each instruction was converted into natural language and used as input for model generation. 
The outputs were validated by directly invoking the official \texttt{sde -S} syntax checker from Synopsys Sentaurus without any manual correction.

It is worth noting that the QA-level and code-level evaluations are performed on two different model checkpoints.
The QA benchmark uses \textsc{TcadGPT-P1\&2}, which was fine-tuned solely on 1.5M Alpaca-style QA pairs.
The code-level executability test, by contrast, evaluates the \textsc{TcadGPT-P1\&2+DPO} model,
which further incorporates IR-based preference alignment to enhance instruction compliance and code correctness.
This separation reflects the distinct objectives of the two evaluation tracks:
semantic understanding versus executable code generation.

\paragraph{Evaluation Protocol.}
For each instruction, we collected both single-sample and multi-sample generations:
\begin{itemize}
    \item \textbf{Pass@1}: proportion of first-sample generations that successfully passed syntax validation without modification.
    \item \textbf{Pass@3}: proportion of instructions with at least one syntactically valid script among three sampled generations.
\end{itemize}
We further categorized results into:
(i) \textbf{Direct pass} — code compiled without modification; 
(ii) \textbf{Placeholder-resolved pass} — code containing template variables (e.g., \texttt{@height@}, \texttt{@gate\_len@}) that passed syntax check after substituting numerical constants; and 
(iii) \textbf{Fail} — code rejected by the compiler due to syntax or structural errors.
All tests were conducted using the same SDE version and configuration to ensure consistency.

\begin{table}[h]
\centering
\caption{%
Syntax-level executability on a 20-instruction SDE test set. 
Pass@k: proportion of instructions whose generated code passed \texttt{sde -S}. 
Placeholder resolution refers to replacing template variables (e.g., \texttt{@height@}) with numeric constants.
}
\label{tab:sde_exec}
\small
\resizebox{\columnwidth}{!}{%
\begin{tabular}{lcc}
\toprule
\textbf{Model} & \textbf{Pass@1} & \textbf{Pass@3} \\
\midrule
\textsc{TcadGPT (P1+P2+DPO)} & 13/20 (65.0\%) & 16/20 (80.0\%) \\
\textsc{DeepSeek V3}          & ~~0/20 (0.0\%) & ~~0/20 (0.0\%) \\
\bottomrule
\end{tabular}%
}
\vspace{3pt}
\raggedright\footnotesize
\textit{Breakdown for \textsc{TcadGPT}:} 
Direct-pass@1 = 12/20 (60.0\%); after placeholder resolution = +1 $\Rightarrow$ Pass@1 = 13/20. 
Direct-pass@3 = 15/20 (75.0\%); after placeholder resolution = +1 $\Rightarrow$ Pass@3 = 16/20.
\end{table}

\begin{figure}[t]
    \centering
    \begin{subfigure}[t]{0.48\linewidth}
        \centering
        \includegraphics[width=\linewidth]{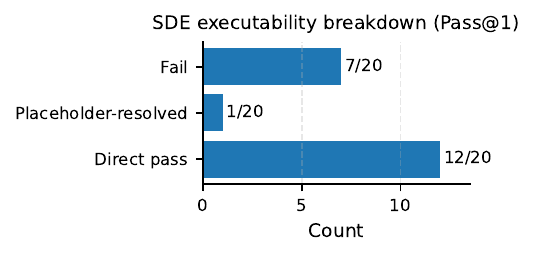}
        \caption{}
        \label{fig:sde_exec_pass1}
    \end{subfigure}
    \hfill
    \begin{subfigure}[t]{0.48\linewidth}
        \centering
        \includegraphics[width=\linewidth]{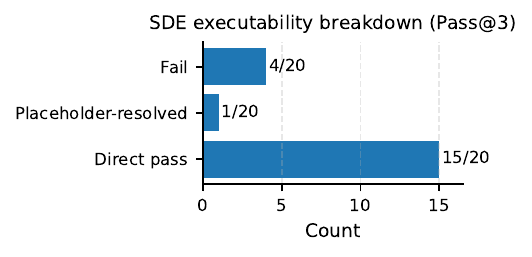}
        \caption{}
        \label{fig:sde_exec_pass3}
    \end{subfigure}
    \caption{%
    \textbf{SDE executability breakdown for \textsc{TcadGPT}.}
    (a) Pass@1 results showing direct passes, placeholder-resolved passes, and failures.
    (b) Pass@3 results under three-sample decoding.
    }
    \label{fig:sde_exec_breakdown}
\end{figure}

\paragraph{Results and Discussion.}
As summarized in Table~\ref{tab:sde_exec} and Fig.~\ref{fig:sde_exec_breakdown}, 
the final \textsc{TcadGPT} model—fine-tuned on the combined \textbf{P1+P2 QA corpus and the DPO alignment dataset}—achieved a \textbf{Pass@1 rate of 65.0\%} (13 out of 20 instructions) on the syntax-level SDE executability benchmark.
As shown in Fig.~\ref{fig:sde_exec_breakdown}(a), the majority of successful cases were \emph{direct passes} without any modification (12/20), with only a single additional case passing after simple placeholder substitution.

When sampling three generations per instruction (\textbf{Pass@3}), the overall pass rate further increased to \textbf{80.0\%} (16/20).
Importantly, as illustrated in Fig.~\ref{fig:sde_exec_breakdown}(b), this improvement was primarily driven by an increase in direct passes (15/20), rather than reliance on placeholder resolution, indicating improved structural robustness under multi-sample decoding.

In contrast, \textsc{DeepSeek V3}—despite producing outputs that superficially resembled SDE scripts—failed syntax validation for all 20 test cases (\textbf{0\% Pass@1 / Pass@3}), confirming that its generated code lacked structural correctness and command-level consistency.

\paragraph{SDE Test Set.}
The 20-instruction SDE test set was designed to comprehensively evaluate the executability of model-generated scripts across representative TCAD scenarios. 
It covers both 2D and 3D structures, diverse Boolean operation modes (ABA, BAB), material stacks (Silicon, SiO$_2$, Si$_3$N$_4$, PolySi, GaN, AlGaN, etc.), and a range of procedural commands including contact definition, doping placement (constant and Gaussian), mesh refinement, and export directives. 
Instructions were automatically rendered into natural language via the IR generator, ensuring that they are unseen during training while remaining fully valid within the Synopsys SDE syntax space.
This mixture of geometry, doping, and meshing tasks provides a practical and challenging benchmark for assessing \emph{syntax-level executability} in real TCAD workflows.

\paragraph{Why DPO is Necessary.}
Before adopting Direct Preference Optimization (DPO), we systematically explored several alignment strategies based on standard supervised fine-tuning. 
Models trained solely on the Alpaca-style QA corpus, even after introducing chain-of-thought (COT) rationales, exhibited poor instruction adherence and frequently produced \emph{almost-correct but non-executable} code: 
commands were placed in the wrong order, mesh parameters were mis-scaled, and export directives were often omitted. 
These issues persisted regardless of prompt format or data scale, revealing that simple SFT or COT supervision was insufficient to instill strict syntactic discipline. 
Integrating DPO on top of the QA-tuned base proved essential, as it directly optimized the model to prefer instruction-compliant outputs while penalizing numeric or procedural deviations. 
This shift led to a clear and reproducible improvement in executable accuracy—evident in the 80\% syntax pass rate achieved in the SDE test set.

These results highlight that integrating DPO alignment on top of QA fine-tuning (P1+P2) significantly enhances instruction compliance and syntactic fidelity. 
Whereas QA-only fine-tuning improves general reasoning but not formal syntax, the DPO stage explicitly optimizes for executable consistency by penalizing numeric and procedural deviations (e.g., wrong mesh parameters, export omissions). 
Consequently, the resulting model can interpret simulation intent and generate scripts that are \emph{directly compilable by commercial TCAD tools}.

\paragraph{Scope and Limitations.}
This evaluation focuses on \textbf{syntax-level executability} in the SDE environment.
Runtime validation (e.g., full mesh build and TDR/BND export) was not included.
The sample size (20 instructions) is limited but representative of diverse geometry and doping scenarios.
Future work will expand this evaluation to SProcess and SDevice modules, increase coverage to over 100 instructions per tool, and introduce an \textbf{end-to-end execution success (Pass@k)} metric capturing solver-completed runs under a fixed environment.

\paragraph{Summary.}
The code-level experiment shows that combining QA-based fine-tuning (P1+P2) with IR$\rightarrow$DPO alignment substantially improves \emph{instruction-following fidelity} and \emph{syntax-level compilability} in a commercial TCAD toolchain.
With an \textbf{80\%} \texttt{sde -S} pass rate, \textsc{TcadGPT} moves beyond descriptive assistance and becomes a practical generator of \emph{tool-accepted} SDE scripts, whereas the general-purpose \textsc{DeepSeek V3} baseline fails all basic syntax checks in this setting.

\subsection{Lightweight Cross-Domain Demonstration on Elmer}
\label{sec:elmer}

\begin{figure}[t]
    \centering
    \begin{subfigure}[t]{0.48\linewidth}
        \centering
        \includegraphics[width=\linewidth]{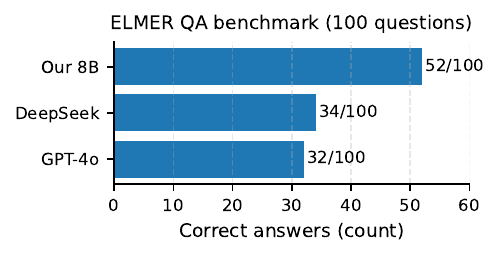}
        \caption{QA accuracy on 100 questions (Pass@1).}
        \label{fig:elmer_qa}
    \end{subfigure}
    \hfill
    \begin{subfigure}[t]{0.48\linewidth}
        \centering
        \includegraphics[width=\linewidth]{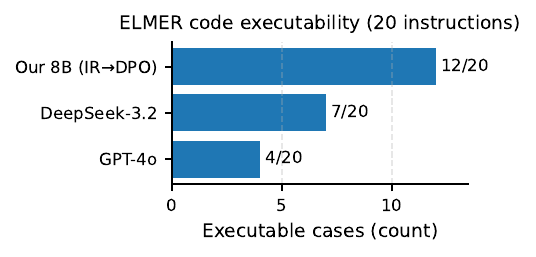}
        \caption{Executable cases on 20 instructions (Pass@1).}
        \label{fig:elmer_code}
    \end{subfigure}
    \caption{\textbf{Lightweight cross-domain check on Elmer.} 
    Using the same QA synthesis and IR$\rightarrow$DPO alignment recipe, our 8B model outperforms general-purpose LLMs on both QA and solver-level code executability.}
    \label{fig:elmer_overview}
\end{figure}

\subsubsection{Setup}
To assess the portability of our schema-first alignment framework beyond TCAD, we conduct a lightweight cross-domain study on the open-source finite-element solver \textit{Elmer}.
Unlike TCAD tools, which are specialized semiconductor simulators, \textit{Elmer} is a general-purpose multiphysics FEM solver with a fundamentally different problem formulation, numerical workflow, and scripting interface.
This structural difference makes it a suitable testbed for probing cross-domain generalization rather than near-domain transfer.

We intentionally keep this study small-scale and complementary.
The goal is not to establish a benchmark comparable in depth to the TCAD suite, but to verify whether the proposed alignment recipe can be instantiated with minimal domain-specific engineering effort.

We reuse the same two-stage recipe as in the TCAD setting:
(i) QA synthesis from domain documentation using Pipeline~2, followed by full-parameter fine-tuning of an 8B base model; and
(ii) an IR$\rightarrow$DPO-style alignment workflow to improve instruction compliance and script executability.
All QA training data are generated automatically from publicly available Elmer documentation, without access to proprietary corpora.

We note that the DeepSeek baseline used in the Elmer experiments corresponds to DeepSeek V3.2, whereas the TCAD experiments use DeepSeek V3.
This difference arises from changes in the official API during the course of the study and does not affect the qualitative comparison reported here.

\subsubsection{QA evaluation}
For QA-level evaluation, we manually construct a 100-question Elmer test set derived from the same documentation sources.
The questions are intentionally restricted to \emph{single-fact} items, each with a single unambiguous answer, covering solver configuration, boundary conditions, material specification, and common scripting patterns.
This design avoids multi-step reasoning and isolates factual recall and command-level understanding.

As summarized in Fig.~\ref{fig:elmer_qa}, under Pass@1 evaluation the Elmer-adapted 8B model achieves \textbf{52/100}, compared with \textbf{34/100} for DeepSeek V3.2 and \textbf{32/100} for GPT-4o.
The consistent margin indicates that domain-specific QA synthesis alone already yields substantial gains over general-purpose LLMs in a structurally distinct solver domain.

\subsubsection{Code executability evaluation}
We further evaluate code-level generation on 20 unseen natural-language instructions.
For each instruction, the model generates a \texttt{.sif/.jis}-style Elmer solver input, which is executed directly using \texttt{ElmerSolver}, without relying on a separate syntax-only checker.

Before execution, we automatically strip any non-\texttt{.sif} content (e.g., explanations or Markdown) from model outputs and ensure that referenced mesh files exist, either by correcting mesh database pointers or by generating meshes via \texttt{ElmerGrid}.
No manual modification of solver commands or numerical parameters is performed.

As shown in Fig.~\ref{fig:elmer_code}, the IR$\rightarrow$DPO-aligned 8B model yields \textbf{12/20} executable cases under Pass@1, compared with \textbf{7/20} for DeepSeek V3.2 and \textbf{4/20} for GPT-4o.
Despite the small scale, this executable gap mirrors the QA trend in Fig.~\ref{fig:elmer_overview} and provides a consistent signal that the proposed alignment recipe transfers beyond TCAD to solver-level code generation in a different numerical and scripting paradigm.

We note that executability rates between TCAD and Elmer are not directly comparable.
TCAD scripts exhibit tighter syntactic and numerical coupling with stricter front-end validation, such that minor deviations can lead to early rejection during parsing or preprocessing.
In contrast, Elmer adopts a more permissive key--value style input and defers some consistency checks to later solver stages.
Therefore, absolute executability levels reflect not only model capability but also inherent differences in language strictness and tool-chain behavior.
The Elmer study is thus intended to probe portability rather than to normalize difficulty across domains.

\section{Conclusion and Future Work}

We presented a \textbf{schema-first alignment framework} for building \emph{executable} domain-specific LLMs under \emph{data scarcity}. 
Instantiated on TCAD as \textsc{TcadGPT}, the framework combines large-scale QA synthesis (1.5M pairs) with IR-driven DPO alignment to directly optimize instruction-following fidelity and syntactic validity for tool-accepted code generation. 
On a 264-question benchmark and a 20-instruction SDE executability test, \textsc{TcadGPT} achieves \textbf{85.6\%} semantic accuracy and an \textbf{80.0\%} syntax pass rate under the official \texttt{sde -S} checker, substantially outperforming general-purpose LLMs, which exhibit markedly lower success rates in this setting. 
We further observe that Retrieval-Augmented Generation (RAG) substantially benefits generic models but can marginally degrade already specialized compact ones, motivating the need for \emph{schema-constrained} retrieval strategies.

\textbf{Beyond TCAD}, we show that the same alignment recipe—QA synthesis from expert materials, schema-first IR extraction and diversification, and IR$\rightarrow$DPO alignment—can be applied with minimal modification to an open-source FEM solver, yielding consistent improvements in script-level success rates. 
These results suggest that the proposed framework captures a reusable alignment pattern for tool-executable LLMs in data-scarce scientific and engineering domains that demand strict artifact validity.

\paragraph{Limitations.}
Our evaluation deliberately emphasizes \textbf{syntax-level executability} (tool acceptance) rather than \textbf{runtime-level} outcomes such as numerical convergence or end-to-end completion. 
The latter can depend sensitively on simulator-specific stability, meshing strategies, and solver configurations, and are therefore not fully isolated by the current benchmarks. 
In addition, the present system does not yet perform robust multi-step planning or interactive repair, and residual hallucinations may occur under weakly structured retrieval settings.

\paragraph{Future Work.}
Future directions include: (i) \textbf{closed-loop integration} with commercial solvers to enable runtime-level verification and automatic repair; 
(ii) \textbf{reward modeling} based on convergence signals, physical consistency checks, or explicit tool feedback; 
(iii) stronger \textbf{planning and decomposition} mechanisms for multi-stage workflows; 
and (iv) \textbf{schema-aware retrieval} that constrains evidence injection to avoid lexical overfitting while preserving factual grounding.


\appendix

\section*{Appendix A: Prompt Templates for Pipeline 1}

\subsection*{Pipeline 1: Prompt for Generating Alpaca-format QA Pairs}

\begin{tcolorbox}[colback=white, colframe=gray!60!black, title=Prompt: Generate QA Pairs]
\small
Your task is to generate high-quality fine-tuning data based on the provided text. In \texttt{content}, you will be given a domain-specific technical passage. You need to generate data in the following format:

\texttt{\{ "instruction": "User instruction.", "input": "", "output": "Expected system output." \}}

Questions must be relevant to the material, professional, and challenging. Focus on commands, code, formulas, operation steps, scientific principles, and software logic. Questions should be diverse, and answers must be rich, professional, and structured.

The output should be a normal response to your generated instruction. Ensure the instruction is general and not limited to phrases like \texttt{in this book}. The format must be strictly followed. The provided content may contain HTML tags or unrendered LaTeX code—only output human-readable text, and ensure formulas render correctly.

Content is mostly technical manuals or textbooks. Ask a question about each technical point or command. Sample questions include:

\begin{itemize}
    \item What is the impact of optical phonon scattering on mobility?
    \item How to define new dopant types in Sprocess?
    \item How to save a simulated structure as a TDR file?
    \item What are the key parameters in ion implantation, and how to configure simulation commands?
\end{itemize}

Avoid duplicate or equivalent questions. Always generate at least 10 different questions per passage. All content must be in Chinese.
\end{tcolorbox}

\subsection*{Pipeline 1: Prompt for Question Augmentation}

\begin{tcolorbox}[colback=white, colframe=gray!60!black, title=Prompt: Augment Questions]
\small
You will be given a single QA pair from a fine-tuning dataset. Your task is to perform data augmentation. Modify the question into different forms.

Output should be a Python list containing 10 new questions.

Ensure question diversity. Questions should be inspired by both the original question and the answer. Output must be a Python-readable list:

\texttt{["question1", "question2", ..., "question10"]}

Only output the list; no extra characters. Must be in Chinese.
\end{tcolorbox}

\section*{Appendix B: Prompt Templates for Keyword-Guided Pipeline 2}

\begin{tcolorbox}[colback=white, colframe=gray!60!black, title=Prompt: Extract Keywords]
\small
You will be given a technical TCAD passage in either English or Chinese. Your task is to extract all keywords that can be further queried. These include but are not limited to:
\begin{itemize}
    \item Command names (e.g., \texttt{pdbSet}, \texttt{Electrode})
    \item Parameter names (e.g., \texttt{MinGrowthStep})
    \item Model names (e.g., Hydrodynamic model)
    \item Software module names (e.g., Sentaurus Device)
    \item Physical mechanisms or equations (e.g., thermionic emission, carrier drift)
\end{itemize}
Output format must be:

\texttt{\{ "keywords": ["keyword1", "keyword2", ...] \}}

Strictly follow this format. Do not extract keywords if the passage lacks technical content. Preserve the original language of each keyword and ensure complete coverage.
\end{tcolorbox}

\vspace{0.5cm}

\begin{tcolorbox}[colback=white, colframe=gray!60!black, title=Prompt: Generate QA from Keywords]
\small
Your task: Based on the given technical paragraph and keyword, generate structured Chinese QA data for Alpaca-style fine-tuning.

Output must be a JSON array. Each entry follows this format:

\texttt{\{ "instruction": "User question", "input": "", "output": "System answer" \}}

\textbf{Guidelines:}
\begin{enumerate}
    \item Only generate in Chinese.
    \item If the paragraph lacks semantic sentences or relevant info for a keyword, skip it and return \texttt{[]}.
    \item Style based on keyword type: concept, formula, command, module, or configuration.
    \item Cover all aspects of the keyword. Avoid repetition.
    \item Answers must be precise, executable, with code or formulas if needed.
    \item Use \texttt{\$...\$} for inline formulas.
    \item Do not refer directly to the paragraph. Do not add markdown markers.
    \item When the keyword is a command/parameter, the question should describe the function, not the name.
\end{enumerate}
\end{tcolorbox}

\end{document}